\documentclass[conference]{IEEEtran}
\IEEEoverridecommandlockouts
\usepackage{cite}				
\usepackage{graphicx}	
\usepackage{amsmath}		
\usepackage{algorithmic}	
\usepackage{array}		%
\ifCLASSOPTIONcompsoc	
\usepackage[caption=false,font=normalsize,labelfont=sf,textfont=sf]{subfig}
\else
\usepackage[caption=false,font=footnotesize]{subfig}
\fi
\usepackage{stfloats}
\usepackage{url}
\usepackage{booktabs}	
\usepackage{algorithm}	
\usepackage{color}		
\usepackage{bm}			
\usepackage{amsfonts}	
\usepackage{amssymb}	
\usepackage{hyperref}	
\usepackage[switch]{lineno}	

\newtheorem{theorem}{Theorem}

\begin{document}

\title{Performance Analysis for Cache-enabled Cellular Networks with Cooperative Transmission}

\author{\IEEEauthorblockN{Tianming Feng\IEEEauthorrefmark{1},
Shuo~Shi\IEEEauthorrefmark{1}\IEEEauthorrefmark{2},
Shushi~Gu\IEEEauthorrefmark{3}\IEEEauthorrefmark{2}, 
Ning~Zhang\IEEEauthorrefmark{4},
Wei~Xiang\IEEEauthorrefmark{5}\IEEEauthorrefmark{2} and
Xuemai~Gu\IEEEauthorrefmark{1}\IEEEauthorrefmark{2}}
\IEEEauthorblockA{\IEEEauthorrefmark{1}School of Electronics and Information Engineering,
			Harbin Institute of Technology, Harbin 150001, China}
\IEEEauthorblockA{\IEEEauthorrefmark{2}Peng Cheng Laboratory, Shenzhen 518055, China}
\IEEEauthorblockA{\IEEEauthorrefmark{3}School of Electronics and Information Engineering, Harbin Institute of Technology (Shenzhen), Shenzhen 518055, China}
\IEEEauthorblockA{\IEEEauthorrefmark{4}Department of Electrical and Computer Engineering,
	University of Windsor, Windsor, ON N9B 3P4, Canada}
\IEEEauthorblockA{\IEEEauthorrefmark{5}College of Science and Engineering, James~Cook~University, Cairns, QLD~4878, Australia}
E-mail: \{fengtianming, crcss, gushushi\}@hit.edu.cn, ning.zhang@ieee.org, wei.xiang@jcu.edu.au, guxuemai@hit.edu.cn
\thanks{This work is supported by the National Natural Sciences Foundation of China under Grant 61701136 and the project ``the Verification Platform of Multi-tier Coverage Communication Network for Oceans (PCL2018KP002)''.}
}


\maketitle
\begin{abstract}
	The large amount of deployed smart devices put tremendous traffic pressure on networks. Caching at the edge has been widely studied as a promising technique to solve this problem. To further improve the successful transmission probability (STP) of cache-enabled cellular networks (CEN), we combine the cooperative transmission technique with CEN and propose a novel transmission scheme. Local channel state information (CSI) is introduced at each cooperative base station (BS) to enhance the strength of the signal received by the user. A tight approximation for the STP of this scheme is derived using tools from stochastic geometry. The optimal content placement strategy of this scheme is obtained using a numerical method to maximize the STP. Simulation results demonstrate the optimal strategy achieves significant gains in STP over several comparative baselines with the proposed scheme.
\end{abstract}
\IEEEpeerreviewmaketitle

\section{Introduction}\label{sec1}

The dense deployment of smart devices results in a tremendous increase in wireless data traffic. Based on the fact that a large portion of the traffic is caused by repeatedly downloading a few popular contents,  caching technique is introduced into cellular networks to relieve the traffic burden, thereby forming cache-enabled cellular networks (CEN). There have been extensive studies on various caching strategies considering different network paradigms such as heterogeneous cellular networks (HetNets) \cite{015Cui2016AnalysisAO, Kuang2019}, device-to-device networks (D2D) \cite{Chen2018,Deng2018}, and IoT networks \cite{Duan2018,Zhao2019}. Specifically, the authors in \cite{015Cui2016AnalysisAO} design an optimal random caching strategy with multicasting in large-scale HetNets to maximize the successful transmission probability (STP). In \cite{Kuang2019}, the authors consider the analysis and optimization of random caching in the $K$-tier multi-antenna multi-user HetNets and obtain a locally optimal solution. As for the cache-enabled D2D networks, \cite{Chen2018} introduce a prior knowledge-based learning algorithm to optimize the caching policy with the knowledge of user preference and activity level to maximize offloading probability. The work in \cite{Deng2018} takes the user mobility into consideration in D2D networks and optimizes caching placement to minimize the cost of obtaining files by user. In the CEN, \cite{Duan2018} investigates the optimal cache space proportion to reduce the average energy consumption of the system, and \cite{Zhao2019} proposes a heuristic routing protocol for CEN with diverse connectivity. 

On the other hand, carefully designing a transmission strategy can further improve the STP of CEN. From this aspect, joint transmission (JT), as one of the base station (BS) cooperation techniques where the user equipment (UE) is served cooperatively by multiple BSs, can be adopted in cache-enabled networks. In \cite{11Wen2017RandomCB}, the authors propose two BS cooperative transmission policies in cache-enabled HetNets, and design an optimal content probability to maximize the STP under each scheme. In \cite{021Chen2016CooperativeCA}, the caching storage is divided into two portions, one of which stores the most popular contents, and the other one cooperatively stores less popular contents in different BSs. \cite{Chae2017} studies the tradeoff between the content diversity gain and the cooperative gain and proposes an optimal caching strategy to balance the tradeoff. However, the aforementioned works just consider the non-conherent JT, where the channel state information (CSI) is not available.

In this paper, we investigate the benefits of CSI in CEN when considering the JT. Different from the conventional JT scheme that shares the global CSI among the cooperative  BSs and precodes the required data with the globally shared CSI, which can further increase the burden on the networks, we propose a novel local CSI based joint transmission (LC-JT) scheme for the CEN. In LC-JT, each cooperative  BS only has the knowledge of CSI of links between itself and its associated UEs, which we refer to as local CSI, and do not share the local CSI among cooperative  BSs. With LC-JT scheme, we analyze the performance of CEN. To this end, we first derive a tight approximation of the STP with LC-JT scheme in CEN, since it is difficult to obtain a closed-form expression for the STP. Then, we verify the tightness of the approximation using simulations and find the optimal content placement strategy using a numerical method to maximize the STP. Finally, simulation results demonstrate the optimal strategy achieves significant gains in STP over several comparative baselines with LC-JT scheme.

\section{System Model and LC-JT Scheme}\label{section: Problem statement}
\subsection{Model of Cache-enabled Cellular Network}
\begin{figure}[!t]	
	\centering
	\includegraphics[scale=0.4]{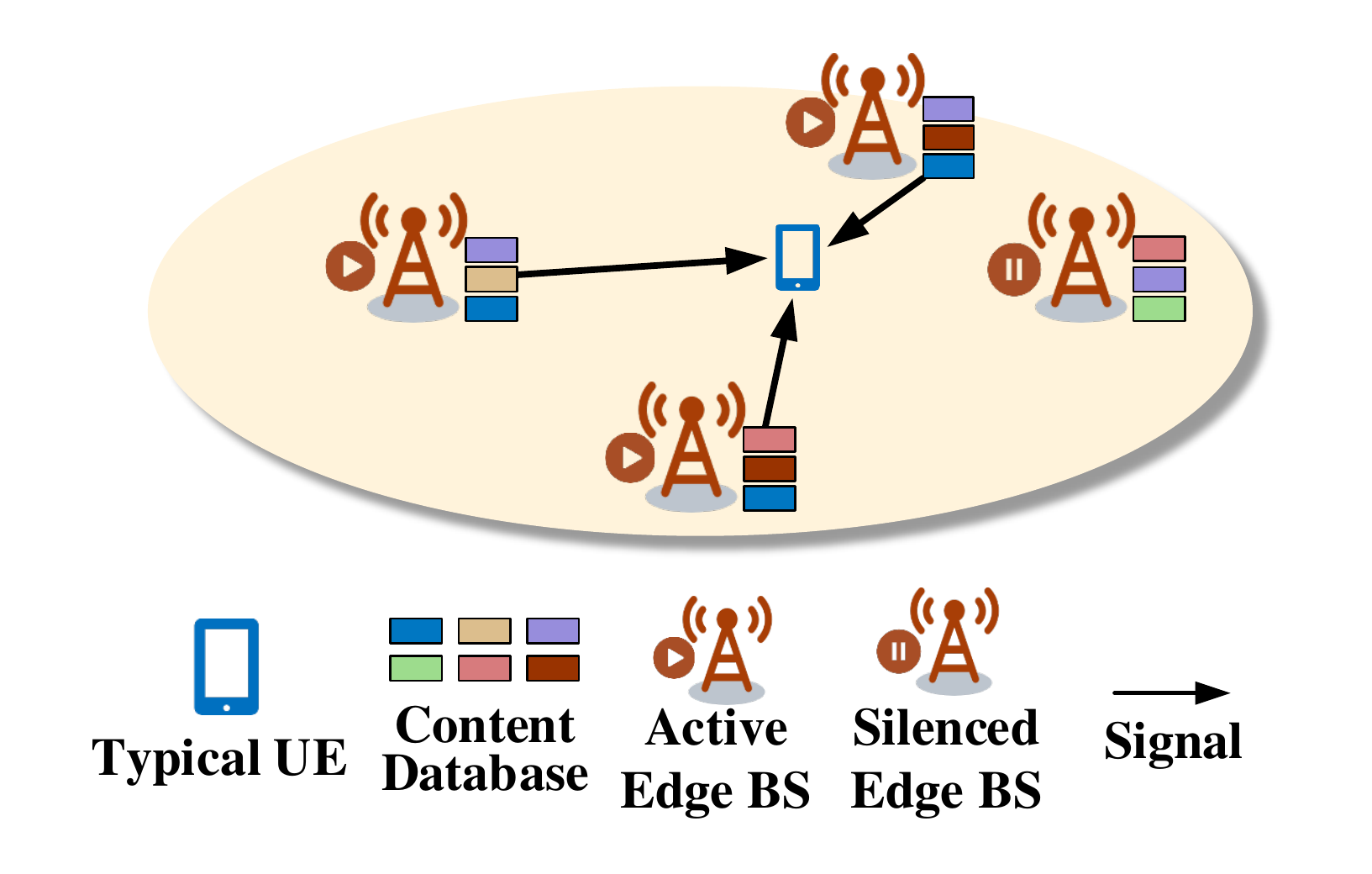} 
	\caption{\textit{Illustration of a CEN with LC-JT scheme. There are six different files ($N=6$) indicated in six different colors in the network. The color of the typical UE represents the file it requests. In this scenario, $M = 4$, $K = 3$, $C_n = 3$.}}
	\label{fig: System Picture}
\end{figure}

We consider a downlink cellular network, where the contents requested by UEs are jointly transmitted by several  BSs, as shown in Fig. \ref{fig: System Picture}. The locations of the  BSs are modeled as a homogeneous PPP $\Phi_{b}$ with density $\lambda_{b}$. The transmission power of each  BS and pathloss exponent are $P_{b}$ and $\alpha>2$, respectively. According to Slivnyak's theorem \cite{01Haenggi2012StochasticGF}, we focus on a typical UE $u_0$ located at the origin. Time is divided into equal-duration time slots, and we just study one slot of the transmission. In this paper, we consider a content database containing $N \geq 1$ files, which is denoted by $\mathcal{N} = \{1,2,\cdots,N\}$. Each file $n\in \mathcal{N}$ has its own popularity $a_n \in [0,1]$ so that $\sum_{n \in \mathbb{N}} a_n = 1$. Here, we assume the popularity distribution follows a Zipf distribution \cite{11Wen2017RandomCB, 015Cui2016AnalysisAO}, i.e., $a_n = \frac{n^{-\gamma}}{\sum_{n\in \mathcal{N}} n^{-\gamma}},\, \text{for }\forall n\in \mathcal{N},$
where the parameter $\gamma \geq 0$ is the Zipf exponent, representing the popularity skewness. The lower indexed file has higher popularity, i.e., $a_1 \geq a_2 \geq \cdots \geq a_N $. The file popularity distribution $\bm{a} \triangleq (a_n) _{n \in \mathcal{N}} $ is assumed to be known \textit{a prior} and is identical among all UEs. Each UE randomly requests one file according to $\bm{a}$ in one time slot.

Each  BS is equipped with a limited cache space and can store $K$ different files out of $N$. It is assumed that $K<N$, which is reasonable in practice. In this work, we adopt a probabilistic content placement strategy, in which $K$ different files are randomly chosen to store at each  BS. Denote by $T_n \in [0,1]$ the probability that file $n$ is stored at one  BS, and by $\mathbf{T} \triangleq (T_n)_{n\in\mathcal{N}} $ be the \textit{placement probability vector}, which is identical for all the  BSs in the network. 

Then, we have \cite{015Cui2016AnalysisAO,11Wen2017RandomCB}:
\begin{eqnarray}
&0 \leq T_{n} \leq 1,& \label{equ: constraint of T 1}\\
&\sum_{n \in \mathcal{N}} T_{n} = K.& \label{equ: constraint of T 2}
\end{eqnarray}
The locations of the  BSs storing file $n$ are modeled as a thinned homogeneous PPP $\Phi_{b,n}$ with density $\lambda_{b} T_n$. Thus, we have $\Phi_{b} \triangleq \bigcup_{n \in \mathcal{N}} \Phi_{b,n}$. Similarly, let $\Phi_{b,-n}$ be the set of the  BSs that do not store file $n$, and thus, it is also a homogeneous PPP with density $(1-T_n)\lambda_b$. In addition, we have $ \Phi_{b,n} \bigcup \Phi_{b,-n} = \Phi_{b}$.
\subsection{LC-JT Scheme} \label{Joint Transmission Strategy}
It is assumed that each  BS and UE is equipped with a single antenna, which means only one UE can be served in each frequency. Orthogonal multiple access methods, e.g., FDMA, are adopted to cater to the simultaneous content requests at the  BSs. Consider that UE $u_0$ requests file $n$. The $u_0$'s $M$ nearest  BSs are involved to cooperatively transmit data. The set of these $M$  BSs is denoted by $\mathcal{C}$. Let $\Phi_{b}^{c}$ denote the remaining  BSs that are not in the set $\mathcal{C}$, i.e., $\Phi_{b}^{c} \triangleq \Phi_{b}\backslash \mathcal{C}$, and let $\mathcal{C}_{n} \triangleq \mathcal{C} \cap \Phi_{b,n}$ denote the set of  BSs that store file $n$ in $\mathcal{C}$. $\mathcal{C}_{n}$ is referred to as the \textit{cooperative set}, and $\mathcal{C}_{-n} \triangleq \mathcal{C} \backslash \mathcal{C}_{n}$. Let $C_n=\left|\mathcal{C}_{n}\right|$. Consider a content-centric association \cite{015Cui2016AnalysisAO}, where the serving  BS of a UE must store the requested file of the UE but may not be its geographically nearest BS. Based on this association principle, we propose the following cooperative transmission policy: 1) If $C_n=M$, all the  BSs in $\mathcal{C}$ jointly transmit file $n$ to UE $u_0$; 2) If $C_n\in [1,M)$, the  BSs in set $\mathcal{C}_{n}$ jointly serve UE $u_0$, and the  BSs in set $\mathcal{C}_{-n}$ become silent; and 3) If $C_n= 0$, UE $u_0$ will be associated with the nearest  BS that stores the file, and all the  BSs in $\mathcal{C}$ become silent. In all the cases above, we assume the  BSs in $\Phi_{b}^{c}$ are active (with this assumption, the performance of the typical UE is only an approximation of that of an arbitrary UE, as illustrated in \cite{11Wen2017RandomCB}). In order to obtain first-order insights of content placement probability design, we assume there is no other UE served by the  BSs in $\mathcal{C}_{-n}$, which can lead to an optimistic performance result.
The received signal at the typical UE $u_0$ in a time slot is
\begin{equation} \label{equ: received signal}
\begin{aligned}
y \!=\! \!\!\sum_{x \in \mathcal{C}_{n}}\!\! P_{b}^{1 / 2}\|x\|^{-\alpha / 2} h_{x} w_{x} X\! +\!\! \!\sum_{x \in \Phi_{b}^{c}}\!\! P_{b}^{1 / 2}\|x\|^{-\alpha / 2} h_{x} w_{x} X_{x}\!+\!Z,
\end{aligned}	 
\end{equation}
where $\|x\|^{-\alpha}$ and $\left|h_{x}\right|^2$ correspond to large-scale fading and small-scale fading (i.e., Rayleigh fading $ h_x \stackrel{d} {\sim} \mathcal {CN}(0,1)$) of the link between $u_0$ and the  BS located at $x$, respectively; $X$ denotes the transmitted symbol from the  BSs in the cooperative set $\mathcal{C}_n$; $X_x$ denotes the interfering symbol sent by the  BSs outside $\mathcal{C}$; $Z \stackrel{d}{\sim} \mathcal{CN}\left(0, N_{0}\right)$ represents the background thermal noise; $\omega_x$ denotes the precoder used by the  BS located at $x$. Consider a novel LC-JT scheme, where we assume only the CSI of the links between each  BS and its associated UEs is available at each  BS, which we refer to as \textit{local CSI}. Given local CSI $h_x$, the precoder $\omega_x$ in \eqref{equ: received signal} is designed as 
\begin{equation}
w_{x} = h^{*}_{x}/\left|h_{x}\right|,
\end{equation}
where $h^*_x$ denotes the complex conjugate of $h_x$. Throughout this paper, we assume that all the fading coefficients $h_x$ are i.i.d. The precoded signal is jointly transmitted to the UE by all the cooperative  BSs. Obtaining the local CSI can be done using pilot estimation \cite{21Yin2013ACA}, of which the scope is beyond this paper. We assume perfect local CSI at each edge BS. In this work, we just focus on the interference-limited regime. Then, the signal-to-interference ratio (SIR) of the typical UE $u_0$ requesting file $n$ is given by
\begin{equation} \label{equ: SINR_n}
\text{SIR}_{n}\! = \!\frac{\left|\sum_{x \in \mathcal{C}_{n}} \!P_{b}^{1 / 2}\|x\|^{-\alpha / 2} h_{x} w_{x}\right|^{2}}{\sum_{x \in \Phi_{b}^{c}} \!P_{b}\|x\|^{-\alpha}\left|h_{x} w_{x}\right|^{2}}.
\end{equation}
\subsection{Performance Metric} \label{Performance Metric}
In this paper, we consider the file successful transmission probability, i.e., STP to evaluate the performance of CEN. STP refers to the probability that a file is transmitted successfully from an edge BS to its UE. For a requested file $n$ of UE $u_0$, if the achievable transmission data rate exceeds a target threshold $r$ $[\mathrm{bps/Hz}]$, i.e., $\log_{2}(1 + \text{SIR}_n) > r$, $u_0$ can decode the file correctly. Furthermore, the STP is defined as
\begin{equation}\label{equ: STP defined}
q(\mathbf{T}) \triangleq \operatorname{Pr}\left[\text{SIR} \geq \tau \right] = \sum_{n \in \mathcal{N}} a_{n} q_n(T_n),
\end{equation}
where $\tau = 2^r-1$ denotes the SIR threshold; $a_n$ is the popularity of file $n$; the second equality holds due to the total probability theorem; $q_n(T_n) \triangleq \operatorname{Pr} \left[ \text{SIR}_n \geq \tau \right] $ denotes the conditional STP when $u_0$ requests file $n$.
\begin{figure*}[b]	
	\hrulefill
	\begin{equation} \label{equ: q_n0}
	\begin{aligned}
	q_{n,0}(T_n) = \int_0 ^\infty \!\!\int_0 ^{u_0} \exp \left( -A \left( \tau, u_0 \right) - A \left( \tau \left(\frac{u_0}{u_M}\right) ^{\frac{\alpha}{2}}, u_M \left( \frac{1}{T_n} - 1 \right) \right) \right) \frac{u_M^{M-1}}{\Gamma (M)T_n^M} \mathrm{d}u_M \mathrm{d}u_0,
	\end{aligned}
	\end{equation}		
	\begin{equation}\label{equ: R_1}
	R_{m,1} =
	\left\{
	\begin{array}{ll}{\sum\limits_{j=1}^m(-1)^{(j+1)} \binom{m}{j} \!\int_{0}^{\infty} \int\limits_{\substack{\forall t_i \in [0,1] \\ i = 1,\cdots,m}} \exp\left( - A \left( \frac{j\tau} {\sum_{i=1}^{m} t_i^{-\alpha/2}}, u \right) \right) } { \frac{u^ {M-1}} {\Gamma(M)}\mathrm{d}{t_1}\cdots \mathrm{d}{t_m} \mathrm{d}u,} & {m = 1,2 \cdots, M-1,}
	\\ {0,} & { m=M, }
	\end{array}\right.
	\end{equation}
	\begin{equation} \label{equ: R_2}
	R_{m,2} =
	\left\{
	\begin{array}{ll}{ \int\limits_{0}^{\infty} \exp\left( - A\left( \tau , u \right) \right) \frac{u^{M-1} }{\Gamma(M) } \mathrm{d}u, } & { m=1, }
	\\ { \sum\limits_{j=1}^m(-1)^{(j+1)} \binom{m}{j} \! \int_{0}^{\infty}\!\! \int\limits_{\substack{\forall t_i \in [0,1] \\ i = 1,\cdots,m-1}} \exp \left( -A \left(\frac{j \tau}{ 1+\sum_{i=1}^{m-1} t_i^{-\alpha/2} } ,u \right) \right) \frac{u^{M-1} }{\Gamma(M) } \mathrm{d}{t_1} \cdots \mathrm{d}{t_{m-1}} \mathrm{d}{u},} & {m = 2, \cdots, M.}
	\end{array}\right.
	\end{equation}
	%

\end{figure*}
\section{Approximation on STP of CEN}\label{section: STP Derivation}
In this section, we derive the main results of this work, i.e., the approximation on STP with LC-JT scheme of CEN. Then, we verify the obtained expressions using Monte Carlo simulations.

From the transmission policy with LC-JT introduced in Section \ref{Joint Transmission Strategy}, $q ( \mathbf{T} )$ can be written as
\begin{equation} \label{equ: q^s 2}
q ( \mathbf{T} )\! \!=\!\!\! \sum_{n \in \mathcal{N}}\!\! a_{n}\! \Bigg(\!\! \operatorname{Pr} \!\left[ C_n \!=\! 0\right]\! q_{n,0}(T_n)\! +\!\! \sum_{m=1}^{M}\!\! \operatorname{Pr} \left[ C_n \!= \!m\right]\! q_{c,m} \!\!\Bigg)\!.	
\end{equation}
The items in the parentheses are from the total probability theory, where $q_{n,0}(T_n)$ denotes the conditional STP conditioned on $C_n=0$ when required file is $n$; $q_{c,m}$ denotes the conditional STP conditioned on $C_n = m, m=1,2 \cdots,M$ when required file is $n$, and the corresponding normalized received signal power is $ S = \left| \sum_{i=1}^{m} \| x_i \| ^{-\alpha/2} \left|h_{x_i}\right| \right|^2 $, which is normalized by $P_b$. In the process of calculating $q_{c,m}$, a conditional complementary cumulative density probability (CDF) $ \operatorname{Pr} \left[ S \geq \tau I \left| \right.\bm{R} = \bm{r},I\right] $ needs to be considered, where $\bm{R} = \left( R_{1}, \cdots, R_{m}, R_{M} \right) $, $m = 1,2,\cdots,M-1$; $M\in\mathbb{N}^+$ and $R_i,i\in \mathbb{N}^+$ denotes the distance between the $i$-th nearest edge BS and $u_0$. Since $\left|h_{x}\right|$ is a Rayleigh distributed random variable (RV), $S$ is the square of the weighted sum of Rayleigh RVs, whose CDF still can not be expressed explicitly. However, we can obtain an approximation of $q_{c,m}$.
The approximation of $q ( \mathbf{T} ) $ are given by the following theorem using stochastic geometry.
\begin{theorem}[Approximation of STP]\label{theorem: upper and Appr of C-JT} 
	The approximation of the STP $q( \mathbf{T} ) $ with LC-JT is given by
	\begin{equation}\label{equ: Appr of qCSI}
	q^{a} ( \mathbf{T} ) = \sum_{n \in \mathcal{N}} a_{n} q_n^{a}(T_n),
	\end{equation}
	where $q_n^{a}(T_n)$ is given by
	\begin{equation}\label{equ: qnCSIappro2 in theorem 3}	
	\begin{aligned}
	q_n^{a}(\!T_n\!)\!\! =\!\! \left(\! 1 \!- \!T_{n}\! \right) ^ {\!M}\!\! q_{n,0}(\!T_n\!) \!\!+\!\! \!\sum_{m=1}^{M}\!\! \dbinom{M}{m}\! T_{n}^{m} \!\left(\!1\!-\!T_{n}\!\right) ^{\!M\!-\!m}\!\!q_{c,m}^{a}.
	\end{aligned}	
	\end{equation}
	Here $q_{n,0}(T_n)$ is in \eqref{equ: q_n0},	
	where $\Gamma(\cdot)$ denotes the complete Gamma function and $q_{c,m}^{a} \triangleq \left(1-\frac{m}{M} \right) R_{m,1} + \frac{m}{M} R_{m,2}$. $R_{m,1}$ and $R_{m,2}$ are given by \eqref{equ: R_1} and \eqref{equ: R_2}, respectively. In \eqref{equ: q_n0}, \eqref{equ: R_1}, and \eqref{equ: R_2}, we set $A( \theta, u ) = \frac{2 u \theta}{\alpha-2}F_G(\alpha,\theta)  + u$, where $F_G(\alpha,\theta) \triangleq {_2F_1}\left( 1,1-\frac{2}{\alpha}, 2-\frac{2} {\alpha} ,- \theta \right)$ denotes the Gauss hypergeometric function, and $\alpha$ is the path-loss exponent.
\end{theorem}
\begin{IEEEproof}	
	According to the probabilistic content placement strategy, the probability mass function of $C_n$ follows a binomial distribution with parameters $M$ and $T_n$, i.e., $\operatorname{Pr}\left[C_n=m\right] = \dbinom{M}{m} T_{n}^{m} \left( 1 - T_{n} \right) ^ {M-m}$.
	To calculate $q_{n,0}(T_n)$, we rewrite the interference in \eqref{equ: SINR_n} as $I = I_n+I_{-n}$, where $I_{n} \triangleq \sum_{x \in \Phi_{b,n} \backslash \{x_0\}} \|x\|^{-\alpha} \left|h_{x}\right|^{2} $ and $I_{-n} \triangleq \sum_{x \in \Phi_{b,-n} \backslash \mathcal{C}} \|x\|^{-\alpha} \left|h_{x}\right|^{2} $, and $x_0$ denotes the only serving edge BS of $u_0$, the distance between whom is $R_0$. We have $R_0>R_M$, the normalized received power is $ S=R_0^{-\alpha}\left| h_{x_0} \right|^2 $. In this case, we have
	\begin{equation}\label{equ: appendix qn0}
	\begin{aligned}
	&q_{n,0}(T_n)\! \!= \!\int_0 ^\infty \int_0 ^{r_0} f_{\left.R_0,R_M\right|R_0>R_M} \left( T_n, r_0, r_M \right) 
	\\\times &\!\operatorname{Pr}\!\left[R_0^{-\alpha}\! \left| h_{x_0} \right|^2\!/(I_n\!+\!I_{-n})\! \geq\! \tau\! \left.\! \right| \! R_0 \!= \!r_0, \!R_M \!= \!r_M\! \right] \!\mathrm{d}r_{\!M} \mathrm{d}r_0,
	\end{aligned}
	\end{equation}
	where $f_{\left.R_0,R_M\right|R_0>R_M} \left( T_n, r_0, r_M \right)$ is the conditional joint PDF of $R_0$ and $R_M$ conditioning on $R_0>R_M$. We denote the PDF of the distance of the $i$-th nearest edge BS as $f_i(x,\lambda)$. Then, we have
	\begin{equation}\label{equ: appendix f_R0RM}
	\begin{aligned}
	&f_{\!\!\left.R_0,R_M\right|R_0>R_M}\!\! \left( T_n,\! r_0,\! r_{\!M} \!\right)\!\! =\!\! \frac{ f_1(r_0, \lambda_{b}T_n)\!f_M(r_{\!M}\!, \lambda_{b}(1\! - \!T_n))} {\operatorname{Pr}(R_0>R_M)}
	\\&\!=\!\!\frac{4 (\pi\lambda_{b}) ^{ M+1 } T_{n} r_0 r_M ^{2 M-1} } { \Gamma(M) }\exp\!\left(\!-\!\pi \lambda_{b} T_{n} r_0^{2} \!-\!\pi \lambda_{b} \left( 1\!-\!T_{n} \right)\! r_M^{2} \!\right)\!,
	\end{aligned}
	\end{equation}
	Next, we have
	\begin{equation}
	\begin{aligned}
	\operatorname{Pr}&\left[{R_0^{-\alpha} \left| h_{x_0} \right|^2/(I_n+I_{-n})} \geq \tau \left. \right| R_0 = r_0, R_M = r_M \right]
	\\&= \mathbb{E}_{I_n,I_{-n}}\left[ \operatorname{Pr} \left[ \left| h_{x_0} \right|^2 \geq \tau r_0^{\alpha} \left( I_n + I_{-n} \right) \right] \right]
	\\& \overset{(a)}{=} \mathbb{E}_{I_n,I_{-n}}\left[ \exp \left( - \tau r_0^{\alpha} \left( I_n + I_{-n} \right) \right) \right]
	\\&\overset{(b)}{=} \underbrace{ \mathbb{E}_{I_n}\left[ \exp \left( - s I_n \right) \right] }_{\triangleq \mathcal{L}_{I_{n}} \left. \left( s\right) \right|_{s=\tau r_0^{\alpha}} } \underbrace{\mathbb{E}_{I_{-n}}\left[ \exp \left( - s I_{-n} \right) \right] }_{\triangleq \mathcal{L}_{I_{-n}} \left. \left( s \right) \right|_{s=\tau r_0^{\alpha}} }
	\end{aligned}
	\end{equation}
	where (a) is due to $\left|h_{x_0}\right|^{2}\stackrel{d}{\sim} $ Exp(1)$ $, (b) is due to the independence of the homogeneous PPPs.	
	$\mathcal{L}_{I_{n}} \left( s \right) $ and $\mathcal{L}_{I_{-n}}(s)$ represent the Laplace transforms of the interference $I_{n}$ and $I_{-n}$, respectively, and can be written as follows
	\begin{equation}\label{equ: appendix L_In}
		\mathcal{L}_{I_{n}}\!\! \left( s\right) \left.\!\!\right|_{ s = \tau r_0^{\alpha}}\!=\! \exp\!\left(\! -2 \pi \lambda_{b} T_n \frac{ r_0^2}{\alpha -2} \frac{s}{r _0 ^\alpha} F_G\!\left(\! \alpha, \!-\frac{s}{r _0 ^\alpha} \right)\!\right)
	\end{equation}
	\begin{equation}\label{equ: appendix L_I-n}
		\mathcal{L}_{I_{\!-n}}\!\! \left.\left( s \right) \right|_{ s = \tau r_0^{\alpha}}\!\!=\! \exp\!\left(\!\! -2 \pi \lambda_{b} \left(\! 1 \!- \!T_n\! \right) \!\frac{r_M^2}{\alpha\! -\!2} \frac{s}{r _M ^\alpha} 
	F_{\!G}\!\!\left(\!\!\alpha,\! -\!\frac{s}{r _M ^\alpha} \!\right)\!\! \right).
	\end{equation}
 Substituting \eqref{equ: appendix f_R0RM}, \eqref{equ: appendix L_In} and \eqref{equ: appendix L_I-n} into \eqref{equ: appendix qn0} and using $ u = \pi\lambda_{b}T_n r^2 $, we can obtain $q_{n,0}(T_n)$.
	Next, we calculate $q_{c,m} ^{a} $. Let $x_M$ be the $M$-th nearest edge BS in $\mathcal{C}$. We consider two cases: i) $x_M \notin \mathcal{C}_n$ and ii) $x_M \in \mathcal{C}_n$. Conditioning on $C_n = m $, we have
	\begin{equation}\label{equ: appendix q_cm_noCSI }
	\begin{aligned}
	q_{c\!,m} \!\!& = \!\underbrace { \operatorname{Pr} \left[ S/I \!\geq\! \tau \!\left|\! \right. x_M\! \notin\! \mathcal{C}_{n}, C_n\! =\! m \right] } _{ \triangleq q _{c,m,1}} \operatorname{Pr} \left[ x_M \!\notin\! \mathcal{C}_{n} \!\left|\right. C_n\! = \!m \right]
	\\&+\! \underbrace {\operatorname{Pr}\left[ S/I \!\geq \!\tau \!\left|\! \right. x_M\!\in\! \mathcal{C}_{n}, C_n\! =\! m \right] }_ { \triangleq q _{c,m , 2}}\operatorname{Pr} \left[ x_M \!\in\! \mathcal{C}_{n} \!\left|\right. C_n\! =\! m \right]
	\\& \approx\! q^a_{c,m,1} \!\operatorname{Pr}\! \left[ x_M \!\notin\! \mathcal{C}_{n} \!\left|\!\!\right. C_n\! =\! m \right]\!+\! q^a_{c,m,2}\!\operatorname{Pr} \!\left[ x_M \!\in\! \mathcal{C}_{n} \!\left|\!\!\right. C_n\! =\! m \right] 
	\\ &\triangleq q^a_{c,m}.
	\end{aligned}	
	\end{equation}
	Due to the probabilistic content placement strategy, we have $\operatorname{Pr}\left[ x_M \notin \mathcal{C}_{n} \left|\right. C_n= m \right] =1-\frac{m}{M}$ and $\operatorname{Pr}\left[ x_M \in \mathcal{C}_{n} \left|\right. C_n= m \right] =\frac{m}{M},m=1,2,\cdots,M$. 
	
	As for $q^a_{c,m,1}$, when $m=M$, $x_M\notin \mathcal{C}_{n}$ can not happen, thus, we set $q^a_{c,m,1} = q_{c,m,1}=0$. When $m<M$, we take a condition that $\bm{R} = \bm{r}$. We have:
	
	\begin{equation} \label{equ: appendix qm1RCSI }
	\begin{aligned}
	&q_{c,m,1,\bm{R}}\left(\bm{r}\right)
	\\ &=\!\mathbb{E}_I\! \Bigg[\!\! \operatorname{Pr}\! \Bigg[\!\!\left(\! \sum_{i=1}^{m} r_i ^{-\!\frac{\alpha}{2}}\! \!\left| h_{x_i} \!\right| \!\right)^2\!\!\!\! \geq\! \tau \!I \!\left| \right.\! \bm{R} \!=\! \bm{r}, b_M\!\! \notin\! \mathcal{C}_{n},\! C_n\! = \!m \Bigg] \Bigg]
	\\ & \!\overset{(a)}{\leq}\! \mathbb{E}_I \! \Bigg[\! \!\operatorname{Pr}\! \Bigg[ \!\sum_{i=1}^{m} \!\left| h_{x_i} \right| ^2\! \geq\!\! \frac{\tau I}{\omega} \left| \!\right. \bm{R}\! =\! \bm{r}, b_M \!\!\notin \!\mathcal{C}_{n},\! C_n = m \Bigg] \Bigg]
	\\ &\overset{(b)}{=} \!\mathbb{E}_I\! \left[ 1 - \frac{\gamma \left(m, \frac{\tau I} {\omega} \right)}{\Gamma(m)}\right]
	\!\overset{(c)}{\geq} \!1 \!-\! \mathbb{E}_I \left[ \left(1-e^{- \frac{\tau I} {\omega} }\right)^{m} \!\right]
	\\ & = \sum_{j=1}^{m} (-1)^{j+1} \dbinom{m}{j} \underbrace {\mathbb{E}_I \left[\exp{\left( -j  \frac{\tau I} {\omega} \right) } \right] }_{\triangleq \left. \mathcal{L}_I \left( s\right) \right|_{s= j \frac{\tau}{ \omega }}} \triangleq q^a_{c,m,1,\bm{R}}\left(\bm{r}\right),
	\end{aligned}
	\end{equation}
	here, $\omega = \sum _{i=1} ^{m} r_i ^{-\alpha}\overset{d}{\sim} Gamma(m,1)$; (a) follows from the inequality \cite[Eq. (4)]{Hanif2012}; (b) follows from the CDF of gamma distribution; (c) follows from an upper bound on the incomplete gamma function, i.e.,$\frac{\gamma(m, x)}{\Gamma(m)} \leq \left(1-e^{-x}\right)^{m}$.	Similar to \eqref{equ: appendix L_In}, we have
	\begin{equation}\label{equ: appendix L I2}
	\begin{aligned}
	\mathcal{L}_I \left( s\right) \left.\right|_{s= j \frac{\tau}{ \omega }} = \exp\left( -2\pi\lambda_{b} \frac{r_M^2}{\alpha -2} \frac{s}{r_M^\alpha} F_G \left( \alpha, -\frac{s}{r _M ^\alpha}  \right)\!\! \right).
	\end{aligned}
	\end{equation}
	Then, $q^{a} _{c,m,1}$ can be obtained by removing the condition of $q^{a}_{c,m,1,\bm{R}}\left(\bm{r}\right)$ on $\bm{R} = \bm{r}$, whose joint PDF is $f_{\bm{R}}(\bm{r})=\frac{ 2(\pi\lambda_b)^M } {(M-1)!} r_M^{2M-1} e^{-\pi\lambda_b r_M^2 } \prod_{i=1} ^{m} \frac{2r_i} {r_M^2},$. We have
	\begin{equation} \label{equ: appendix qcm1NoCSI integral}
	\begin{aligned}
	q^{a}_{c,m,1}\!=\! \int\limits_{0}^{\infty}\! \int\limits_{0}^{r_M}\! \cdots\! \int\limits_{0}^{r_M} \!q^{a}_{c,m,1,\bm{R}}\!\left(\bm{r}\right) \!f_{\bm{R}}\left(\bm{r}\right) \mathrm{d}r_1\cdots \mathrm{d}r_m \mathrm{d}r_M 
	\end{aligned}.
	\end{equation}
	By using the changes of variables $u=\pi \lambda_{b} r_M^{2}$ and $t_i = \frac{r_i^2}{r_M^2}$, we can obtain $q^{a}_{c,m,1} =R_{m,1} $ in \eqref{equ: appendix q_cm_noCSI }.
	
	$q^a_{c,m,2}$ can be calculated by following the similar steps of calculating $q^{a} _{c,m,1}$. We omit the details due to page limitations. The tightness of this approximation will be demonstrated by comparing with simulation results in the following part.	
\end{IEEEproof}

\begin{figure}[!t]	
	\centering
	\includegraphics[height=5cm,width=5.8cm]{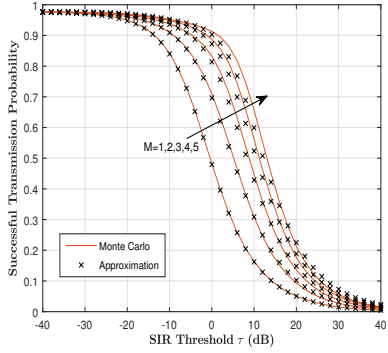} 
	\caption{The approximations and the Monte Carlo results of STP with LC-JT versus $\tau$. 
		$N = 8,$ $K=3,$ $\alpha = 4,$ $\lambda_b = 0.01,$ $\mathbf{T} = [0.9,0.8,0.6,0.4,0.2,0.1,0,0],$ and $\gamma=2.$ In the Monte Carlo simulations, the edge BSs are deployed in a square area of $1,000\times 1,000\, \mathrm{m}^2$, and the results are obtained by averaging over $10^5$ independent realizations.}
	\label{fig: UpperAppr of STP CSI}
\end{figure}

Fig. \ref{fig: UpperAppr of STP CSI} plots $ q^{a} ( \mathbf{T} ) $ versus $\tau$ at different $M$ and the Monte Carlo results. As can be seen, $ q^{a} ( \mathbf{T} ) $ and the Monte Carlo results coincide when $M=1$; when $M\geq2$, $ q^{a} ( \mathbf{T} ) $ tightly approximates the Monte Carlo results over the whole range of $\tau$. The optimal $\mathbf{T}$ to maximize the STP can be obtained by 
\begin{equation}\label{equ: problem}
	\begin{aligned}
	\mathbf{T}^\star = &\arg\underset{\mathbf{T}}{\max} \,\, q^{a} ( \mathbf{T} )
	\\& s.t. \,\, \eqref{equ: constraint of T 1}, \eqref{equ: constraint of T 2}.
	\end{aligned}
\end{equation}
It is hard to get a closed-form solution of \eqref{equ: problem} due to the non-convexity of $q^{a} ( \mathbf{T} )$. Therefore, we obtain the optimal solution by using a numerical method.
\section{Numerical results}\label{section: Simulation results}
In this section, we conduct simulations to validate the analytical results and compare the numerically optimal placement strategy with three baseline strategies, i.e., MPC (most popular caching) \cite{22Batug2014CacheenabledSC}, IIDC (i.i.d. caching) \cite{24Bharath2016ALA} and UDC (uniform distribution caching) \cite{23TamoorulHassan2015ModelingAA}. The numerically optimal placement strategy is obtained using the MATLAB function ``fmincon", and the corresponding algorithm is chosen as ``interior-point-method". Note that the three baselines also adopt LC-JT transmission schemes. Unless otherwise specified, we set $\alpha = 4$, $\lambda_{b} = 0.01$, $M=3$, $N=100$, $K=25$, $\tau = 0\, \mathrm{ dB }$, and $\gamma = 0.8$.
\begin{figure}[!t]
	\centering
	\subfloat[ ] {\includegraphics[height=4.0cm,width=4.5cm]{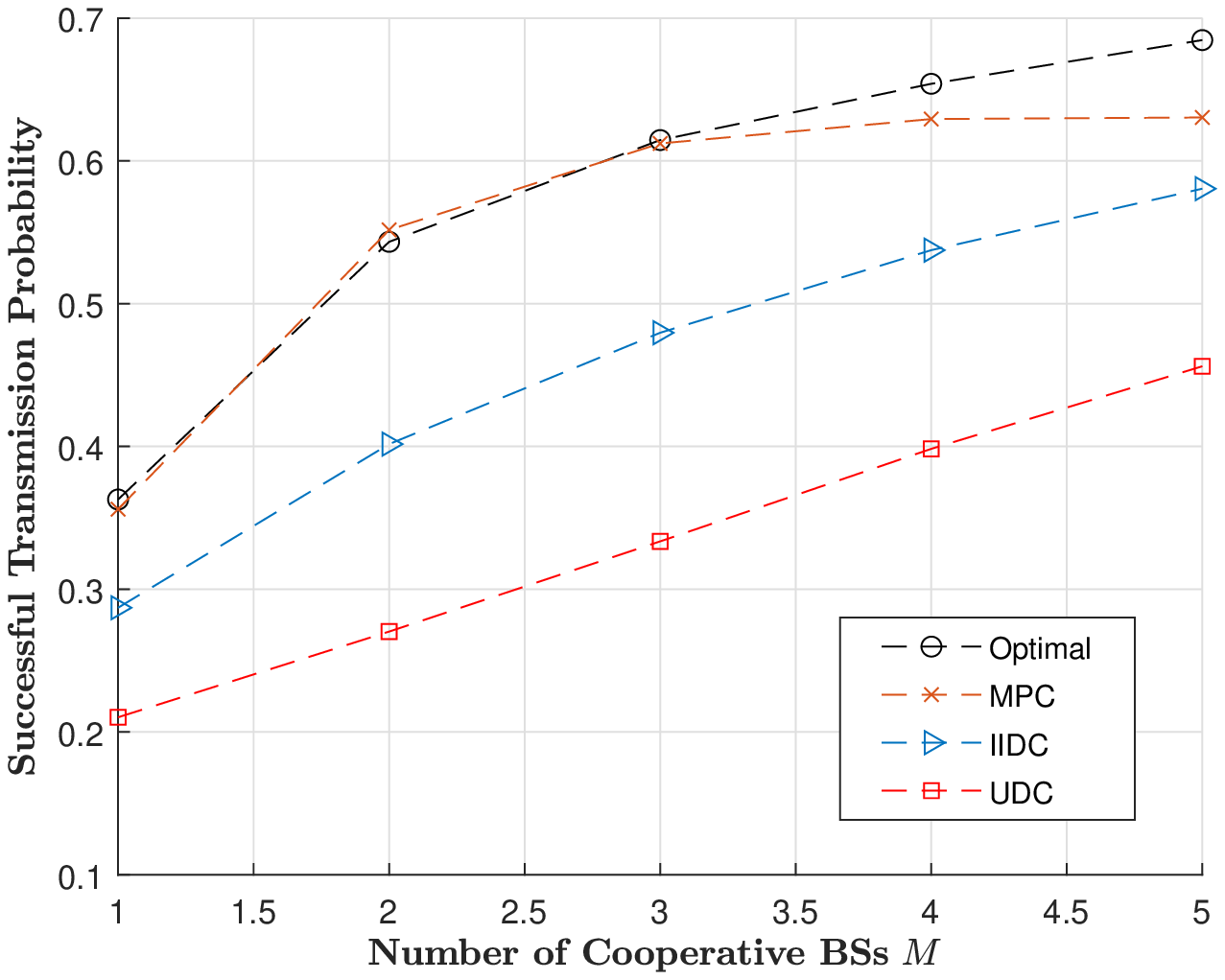}\label{fig: sub STP-M Baseline}}
	\subfloat[] {\includegraphics[height=4.0cm,width=4.5cm]{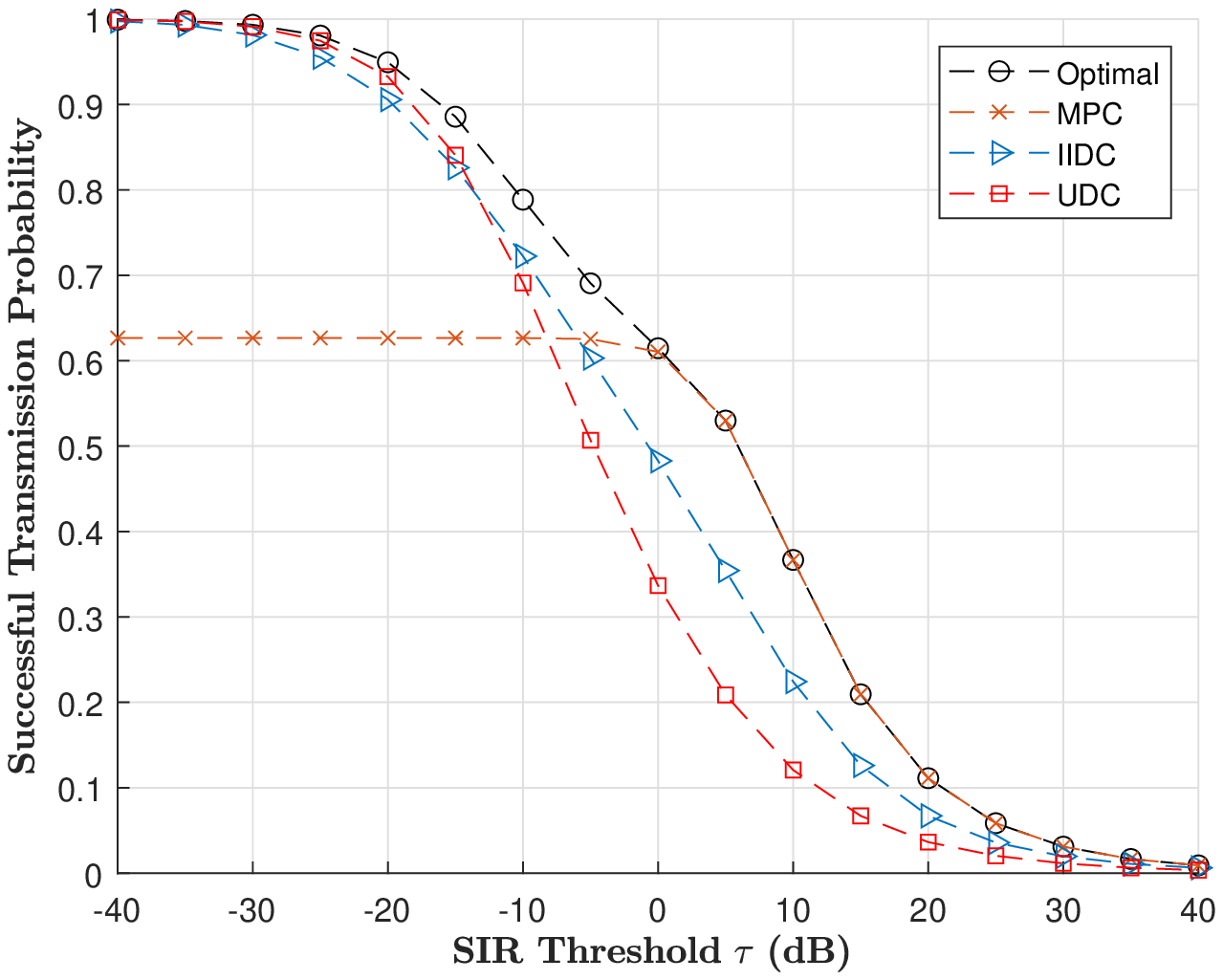}\label{fig: sub STP-tau Baseline}}
	\caption{Comparisons between proposed strategies and baselines with different number of cooperative BSs $M$ and SIR threshold $\tau$ in LC-JT.}
	\label{fig: STP Baseline}
\end{figure}

\begin{figure}[!t]	
	\centering
	\includegraphics[height=5cm,width=5.8cm]{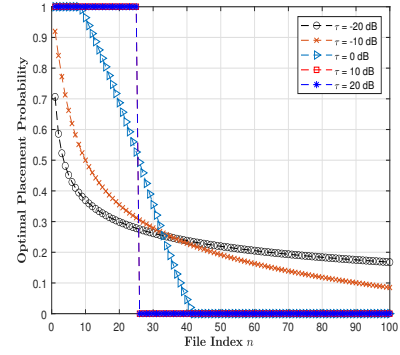} 
	\caption{Optimal placement probability $T_{n}^{\star}$ versus file index $n$ at different SIR thresholds $\tau$.}
	\label{fig: Tn-n tau changes}
\end{figure}

Fig. \ref{fig: STP Baseline} plots the STP versus the number of cooperative BSs $M$, and SIR threshold $\tau$. We can see the optimal placement strategy outperforms all the three baselines when adopting the same transmission scheme. Fig. \ref{fig: STP Baseline}\subref{fig: sub STP-M Baseline} shows that the STPs of all the placement strategies increase with $M$, thanks to joint transmission and BS silencing. As can be seen from Fig.\ref{fig: STP Baseline}\subref{fig: sub STP-tau Baseline}, the STPs of all the placement strategies decrease with $\tau$. In addition, when $\tau>0\,\,\mathrm{ dB }$, the STP of the optimal strategy is the same as that of MPC. Fig. \ref{fig: Tn-n tau changes} depicts the corresponding optimal $\mathbf{T}^\star$. As can be seen, files of higher popularity tend to be stored at BSs. When $\tau$ is high enough, the optimal content placement strategy degenerates to MPC.

\section{Conclusion}\label{section: Conclusion}
In this paper, we proposed a novel local CSI based joint transmission scheme, i.e., LC-JT scheme in CEN. We derived an approximation for the STP of the networks and obtained the optimal content placement strategy numerically. It was shown the approximation is tight and the optimal strategy outperforms several existing baselines.

\bibliographystyle{IEEEtran}
\bibliography{IEEEabrv,VTC2020}

\end{document}